\documentstyle[12pt, epsf]{article}

\begin{document}
\newcommand{\be}{\begin{equation}}
\newcommand{\ee}{\end{equation}}
\newcommand{\lll}{\lambda}
\def\Journal#1#2#3#4{{#1} {\bf #2}, #3 (#4)}

\newcommand{\A}{\alpha}
\newcommand{\B}{\beta}
\newcommand{\T}{\theta}
\newcommand{\Ep}{\epsilon}
\newcommand{\beq}{\begin{equation}}
\newcommand{\eeq}{\end{equation}}
\newcommand{\fr}{\frac}
\newcommand{\beqn}{\begin{eqnarray}}
\newcommand{\eeqn}{\end{eqnarray}}
\newcommand{\G}{\gamma}
\newcommand{\D}{\delta}
\renewcommand{\P}{\phi}
\newcommand{\intl}{\int\limits_{1}^{\infty}}
\renewcommand {\L}{\lambda}
\newcommand{\pt}{\partial}

\newcommand{\bq}{\bar q_A}
\newcommand{\tq}{\tilde q_A}
\newcommand{\btq}{\bar{\tilde q}^A}
\newcommand{\fa}{\varphi^A}
\newcommand{\bfa}{\bar \varphi_A}

\newcommand{\none}{{\cal N}=1}                            
\newcommand{\ntwo}{{\cal N}=2}                            

\def\st{\scriptstyle}
\def\sst{\scriptscriptstyle}
\def\mco{\multicolumn}
\def\epp{\epsilon^{\prime}}
\def\vep{\varepsilon}
\def\ra{\rightarrow}
\def\al{\alpha}
\def\ab{\bar{\alpha}}
\def\bea{\begin{eqnarray}}
\def\eea{\end{eqnarray}}

\begin{flushright}
\begin{tabular}{l}
ITEP-TH-22/08\\
\end{tabular}
\end{flushright}

\vskip1cm
\centerline{\large\bf
Flux-Tube Formation and Holographic  Tunneling}
\vspace{1.5cm}

\centerline{\sc A.S. Gorsky$^{1}$ and V.I. Zakharov$^{1,2}$ }

\vspace{0.5cm}

\centerline{\it 1- Institute of Theoretical and Experimental Physics }
\centerline{\it  B. Cheremushkinskaya ul. 25, 117259 Moscow, Russia}
\centerline{ \it 2- INFN
- Sezione di Pisa,
Largo Pontecorvo, 3, 56127 Pisa, Italy}

\def\thefootnote{\fnsymbol{footnote}}%
\vspace{3.5cm}
\centerline{\bf {Abstract}}

We consider  correlator of two
concentric Wilson loops, a small and large ones related to the problem of
flux-tube formation.  There are three mechanisms which can contribute to the 
connected correlator and yield different dependences on the radius of the small loop. 
The first one is quite standard and concerns exchange by supergravity
modes. We also consider a novel mechanism when the
flux-tube formation is described by a barrier transition in the
string language, dual to the  field-theoretic formulation of
Yang-Mills theories. The most interesting possibility 
within this approach is 
resonant tunneling which would enhance the correlator of the Wilson
loops for particular geometries. 
The third possibility involves  exchange by a dyonic string
supplied with the string junction.
We introduce also 
t'Hooft and composite dyonic loops as probes of the flux tube.
Implications for lattice measurements are briefly discussed.
\vspace{1.0cm}

\newpage
\section{Introduction}
The interior of the chromomagnetic string connecting external heavy
quarks is a traditional subject of theoretical studies devoted to
non-perturbative QCD. There are well known phenomena like
logarithmic  broadening of the string which can be derived from
general grounds   \cite{luscher,greensite}. The broadening follows
from the account of quantum effects on the string worldsheet which
provide the logarithmic dependence of the string width on the length
of the string $L$, $L\gg \Lambda_{QCD}^{-1}$, with $L$ playing the
role of an infrared parameter.

However, as far as dependence on the ultraviolet parameter, say the
lattice spacing $a$, is concerned   it is assumed, at least tacitly,
to be absent. True, the perturbative UV/IR mixing is quite common in the
noncommutative gauge theories (for review see \cite{nekrasov}) and
some features of the noncommutative case could be echoed in the
commutative theories as well. However, in the context of the
structure of the confining strings such ideas have never been
pursued in the literature and the results of the lattice
measurements \cite{boyko} came as a surprise. Namely, according to
\cite{boyko} the width of the flux tube between external quarks
depends on the lattice spacing and, if the pattern observed persists
at smaller lattice spacings, it vanishes in the continuum limit of
$a\to 0$.

The width of the flux tube is defined in terms of the correlator
$\langle s(x), W(T,R)\rangle$ where $s(x)$ is the action density and
$W(T,R)$ is the rectangular Wilson loop of space and time extensions
$R$ and $T$, respectively. In the lattice language, the point $'x'$
is represented by a cube of linear size $a$. In the continuum limit,
it is not possible to evaluate the $a$ dependence since the lattice
spacing serves as an ultraviolet cut off.
Moreover, in the context of the problem considered the lattice
spacing $a$ is to be treated rather  as an uncertainty in position
measurements, or the  size of a measuring device \cite{vz}. To imitate
the problem in the continuum limit we will concentrate, therefore,
on the correlator of two concentric round Wilson loops in the adjoint representation,
\beq
\label{correlator} C(R_{2},R_{1})~=~\langle
W_{R_{2}},W_{R_{1}};h=0\rangle_{connected}~. 
\eeq 
with the radii
$R_{2},R_{1}$ such that $R_{2}\gg R_{1}$, and we fix for simplicity
the distance between the circles planes $h=0$. Moreover, since we
are aiming at comparison with the lattice data, the space-time is
Euclidean.

Once one of the radii, $R_{2}$ is of order $\Lambda_{{QCD}}^{-1}$ or
larger, the field theoretic formulation is not of much help. Since long
\cite{luscher} one relies instead on a string picture.
 Thus, we are invited to consider strings in curved space.  
Note that within this framework the correlator (\ref{correlator})
was considered in detail in Ref. \cite{zarembo} (independently of the
problem of evaluating the profile of the flux tube). In particular, 
it was found that a well defined surface with minimal area 
exists only as far as
the ratio $R_{2}/R_{1}$ is not too large. This is 
an example of a "phase transition" first considered
in the stringy setup in Ref. \cite{go}.
 Beyond a critical value of this ratio
the correlator (\ref{correlator}) cannot be evaluated in terms of the minimal
area of surfaces spanned on the both Wilson loops $W_{1},W_{2}$.

Since we are interested in very large values of the ratio $R_{2}/R_{1}$ we are well
above the ``phase transition'' point of Ref. \cite{zarembo}. To
handle with this regime we have to recognize the mechanism which
would provide the connected part in the correlator
(\ref{correlator}). We shall discuss three possibilities. The first
one is quite conventional and was discussed in a different context
in \cite{berenstein}. Namely, one considers propagation of the
lightest supergravity mode in the bulk between two ``cups'' of the
minimal surfaces in the AdS geometry  ending on the small and large
circles. One expands the small Wilson loop into a set of local
operators and evaluates the correlator of the concentric Wilson
loops in terms of the local operators, like in
\cite{berenstein,zarembo01,thorn}

The second possibility we shall also discuss is more tricky. There
is no classical minimal surface in the Euclidean space-time above
the critical ratio $R_{2}/R_{1}$. However, there exists a complex-action
solution to the equations of motion which involves evolution both in
the Minkowski and Euclidean spaces. Such solutions in the complex
time are quite familiar, say, in the cosmology in the context of the
no-boundary Hartle-Hawking wave function. In a more closely related
setup, complex-time solutions have been considered in the problem of
the multiparticle production on the threshold where they serve as a
generating function for the corresponding production amplitudes
\cite{gv93}. What is important is that such complex
solutions  are intimately related to existence of extended objects  and  in
this sense we are dealing with a  purely stringy phenomena. The
evolution of such complex solution involves several regions. First
it goes in the Euclidean space, then at some ``time'' it jumps into
the Minkowski space-time, undergoes  a nontrivial evolution there
and then goes back to the Euclidean space. Let us emphasize that 
the "`time"' variable here corresponds to the RG scale.

The most subtle point concerns the Minkowski part of the evolution
which  in the case considered in  \cite{gv93} is interpreted  in terms
of the semiclassical bubbles of spherical worldsheets. To ensure
stability of such objects the Bohr-Sommerfiedl quantization
condition was imposed on the Minkowski bubbles and it turns out
that account of such semiclassically quantized bubbles
drastically modifies the probability when the boundary conditions in
the Euclidean region correspond to the resonance condition fixed by the
Bohr-Sommerfield quantization in the Minkowski region. Actually this
is an example of the so-called resonant tunneling phenomenon,  well
known in solid-state physics (see also recent discussion 
in \cite{saffin}). 
Its essence is very simple: if there
exist two barriers with a classically allowed region between them
the tunneling through the  barriers becomes unsuppressed if the
energy of the incoming particle coincides with the energy of a
metastable state between two barriers. We shall discuss possible
resonant tunneling phenomena in our case and derive the 
resonance condition.

Another possible mechanism concerns  magnetic or dyonic strings
and their junctions.
One can ask if there exists a minimal surface 
in the Euclidean space-time providing connectness of the 
Wilson loops 
which involves the string junctions. We argue that there
is such a solution involving propagator of the dyonic string
in the bulk. The dyonic string is attached to the small and
large  Wilson loops by junctions.
We will briefly explore more general 
correlators of the Wilson and t'Hooft loops as well as
of the composite loops made of arcs 
of Wilson and t'Hooft lines.

The organization of the paper is as follows. In Section 2 we shall
briefly review  phenomenology of the flux tube. In Section 3 we
consider the possible minimal surfaces in the AdS geometry and
discuss exchange by the supergravity  modes. In Section 4 we consider the
complex-time solutions and discuss the possibility of resonant
tunneling. Section 5 is devoted to the consideration
of the mechanism of connectness  via dyonic string exchange. 
In Section 6 we discuss more
general configurations involving Wilson, t'Hooft  and composite loops.

\section{Overview of the standard flux-tube phenomenology}
For the purpose of orientation, let us make first a few simple
phenomenological remarks on the flux tube.
 On the lattice the intensity of non-Abelian fields inside
 the flux tube is defined as follows:
\begin{eqnarray}\label{one}
\Delta s(x;R,a)~\equiv~\langle s\rangle_{0}~-~\langle s\rangle_{W~}~\equiv~\\
\langle s\rangle_{0}~-~{\langle s(x),~W(T,R)\rangle\over
\langle W(T,R)\rangle}~,\nonumber
\end{eqnarray}
where
$W(T,R)$ is the Wilson line of the (Euclidean) time- and space- extensions 
$T$ and $R$, 
respectively, $s$ is the action density in lattice units,
$$ s~=~a^{4}(G_{\mu\nu}^{a})^{2}~~,$$ $\langle s\rangle_{0}$ is the vacuum
expectation value of the action, $\langle (plaquette)\rangle\equiv
1- \langle s\rangle_0$.

Note that for small lattice spacings the vacuum expectation value of
the gluon condensate is divergent,
$$\langle G^{2}\rangle_{0}~\sim~a^{-4}~~,$$
as a manifestation of the uncertainty principle. On the other hand,
the common expectation for the confining fields inside inside the
tube is that they  are soft,
$$\langle G^{2}\rangle_{soft}~\sim~\Lambda_{{QCD}}^{4}~~.$$
In other words, the difference of the
plaquette values (\ref{one}) is expected to be  vanishing in the limit
$a\to 0$: \beq\label{difference} a^{4}\langle
G^{2}\rangle_{0}~-~a^{4}\langle G^{2}\rangle_{W}~\sim~(a\cdot
\Lambda_{QCD})^{4}~~. \eeq

Note  that $\Delta s(R,a)$ is positive, as guaranteed by the exact sum
rules for the string tension \cite{michael}. Moreover, the
difference (\ref{one}) is usually fitted to a Gaussian form:
\beq\label{height} \Delta
s(x;R,a)~\sim~\exp\{-h^{2}(x)/\delta^{2}(R,a)\}~~, \eeq where $h(x)$
is the distance from the point $x$ to the plane determined by the
Wilson loop, and $\delta$ is the width of the flux tube.   It goes
without saying that the standard picture assumes that the width is
independent on the lattice spacing,
$\delta~\sim~\Lambda_{QCD}^{-1}$, and one usually concentrates on
the infrared, or $R$-dependence of the width
\cite{luscher,greensite}.

On the other hand, recent lattice measurements \cite{boyko} found
out that in fact the width (\ref{height}) depends drastically on the
lattice spacing which corresponds to the dependence
on $a$ different from (\ref{difference}). This observation is in contradiction with the
standard picture and our aim here is to investigate whether
string-related phenomenology could accommodate the observation
\cite{boyko}.

\section{Correlator of concentric Wilson lines. Generalities.}

It is well known
nowadays that the dual description of gauge theories is provided  by
strings in curved space. In particular, in the conformal case  the metric is:
\beq\label{metric}
ds^{2}~=~R^{2}_{AdS}{1\over z^{2}}\big(dx_{i}^{2}~+~dz^{2}\big)~~,
\eeq ..

The expectation value of the Wilson loop is  controlled by the minimal
value of the area of surface spanned on the loop \cite{maldacena}.
One can then readily see that the effective action, or area is given by
\beq\label{action}
S~=~2\pi R_{AdS}^{2}\Big(\int dz {r\over z^{2}}\sqrt{1~+~(r^{'})^{2}}\Big)~~,
\eeq
where the limits of integration over $z$ are set by boundary conditions,
or the shape of the Wilson loops.
The equation of motion, corresponding to
the action (\ref{action}) reads as:
\beq\label{motion}
{d\over dz}{rr^{'}\over z^{2}\sqrt{1+(r^{'})^{2}}}~=~
{\sqrt{1+(r^{'})^{2}}\over z^{2}}~~.
\eeq

The solution for the minimal area spanned on a circular Wilson loop, 
relevant to our case, is well known. 
To derive the solution, it is convenient
to introduce $(\rho,\tau)$ coordinates instead of $(z,r)$, 
where $r$ is the radial
coordinate on the boundary:
\beq\label{coordinates}
z~=~e^{\tau}cosh^{-1}\rho~~~r~=~e^{\tau}tanh\rho~~.
\eeq
This substitution makes the conformal symmetry manifest,
and one can readily find an integral of motion
which exists as a reflection of this symmetry.
Denoting the value of this integral of motion as $c$, one can 
derive an equation
for the minimal surface in terms of this constant $c$:
\beq\label{integral}
\int d\tau~=~\pm c\int
{d\rho\over cosh\rho\sqrt{sinh^{2}\rho~ cosh^{2}\rho-c^{2}}}~~,
\eeq
where the limits of integration are set by the boundary conditions.

For a single circular Wilson loop the minimal surface is given by the
solution  $\tau=const$ so that the left-hand side of Eq. (\ref{integral})
vanishes  and, therefore,  $c=0$.
The corresponding action, or area
is given by
\beq\label{answer}
S_{R}~=~R^{2}_{AdS}[{R\over \epsilon}~-~1]~~,
\eeq
where $\epsilon$ is an ultraviolet cut off needed to regularize
divergence in action (\ref{action}) at $z\to 0$. The divergent term is
proportional to the perimeter of the Wilson line and corresponds to the
self energy of the external W-boson.  

The case of two concentric Wilson loops was studied first in Ref. \cite{zarembo},
see also \cite{hirata}.
For us the central point is that minimal surface can be determined only
if the ratio $R_{2}/R_{1}$ is not too large. Knowing the 
condition (\ref{integral})
allows to readily derive this result.  Indeed, we need now $z\to 0$ both for
$r=R_{1}$ and $r=R_{2}$. Moreover, $z\to 0$ implies $tanh\rho =1$.
Then the left hand side of Eq. (\ref{integral}) is equal 
to $\ln (R_{2}/R_{1})$.
On the other hand the right-hand side of Eq. (\ref{integral}) is
limited by a quantity of order unit, for any $c$. Thus, 
large $\ln (R_{2}/R_{1})$
is not allowed. More details can be found in \cite{zarembo,hirata}.


Let us first consider the mechanism of connectness above
the critical point via  exchange by supergravity modes. That is we
shall discuss the configuration of two cups corresponding to small
and large Wilson loops. They are connected through the exchange by
supergravity modes described by bulk-bulk propagators. In our case, the
calculation can be simplified since one of the Wilson
loops is very small. That is, it is natural to replace it by a local
operator and therefore the bulk-bulk propagator can be replaced by
a bulk-boundary one. Hence we arrive at the problem of calculation
of the correlator of local operator with the large circular Wilson loop. Such a
problem has been discussed before in \cite{berenstein,zarembo01}.

We substitute the small Wilson loop by a sum over local operators $O_N$
with canonical dimensions $\Delta_N$:   \beq W(R_1)=<W(R_1)>\sum_N
c_N(R_1)O_N,\qquad c_N\propto R_1^{\Delta_N} \eeq and select the
operator with the minimal dimension $O=TrG^2$ which couples to the
dilaton mode in the bulk. The mass of the bulk mode coupled to the
operator with dimension $\Delta$ in $AdS_5$ reads as \beq
m^2=\Delta(\Delta-4) \eeq that is the dilaton is a massless
mode in the AdS bulk. The correlator reduces to the product of the
bulk-boundary propagator of the dilaton and of the vacuum expectation
value of the dilaton vertex operator integrated over the string
worldsheet with the fixed boundary condition on the large circle of
radius $R_2$. Since we are interested in the $R_1$ dependence of the
correlator this one-point function on the worldsheet provides the
$R_1$ independent multiplier which could be calculated
similar to \cite{berenstein,zarembo01}. The bulk-boundary propagator
evidently does not depend on $R_1$ as well hence we arrive at the
following form of the correlator \beq <W(R_1),W(R_2)>\propto
\frac{R_1^4}{R_2^4}F(R_2) \eeq with a calculable function
$F(R_2)$.

So far we discussed the conformal case which admits for exact
solutions. Generalizations to the  realistic QCD case require for extra assumptions.  
For the sake of estimates, we
can use a simple model with a phenomenologically
motivated  modification of the metric (\ref{metric}) which
explicitly introduces the $ \Lambda_{QCD}$ scale.

In particular to get qualitative estimates,  one may have in mind, 
for example, the following
modification of the metric (\ref{metric}): \beq\label{metric1}
{1\over z^{2}}~~\to~~{1\over z^{2}}\exp(const \cdot z^{2})~~, \eeq
where $const \sim \Lambda_{QCD}^{2}$, see, e.g. \cite{andreev} and
references therein.

The crucial point is that at short distances the metric
(\ref{metric1}) is still conformal nd one can still rely on the
derivations valid in the conformal space. 
On the other hand, the modification (\ref{metric1})  introduces an
infrared cut off at $R_2\sim \Lambda_{QCD}^{-1}$. Thus, for the sake
of estimates in the realistic case one can rely on the
conformal-case relations provided the following substitution is
made: \beq R_{1}~~\to~~a,~~R_{2}~~\to~~\Lambda_{QCD}^{-1}~~,~~
R_2/R_1~\to~(\Lambda_{QCD}\cdot a)^{-1}~. \eeq With this rule in
hand, one readily derives predictions for the realistic case.

\section{AdS geometry, complex time solutions and resonant tunneling}

 In the absence of the minimal surface with our boundary conditions
in the Euclidean space we could ask if there is a solution to the
equation of motion in the complex ``time''. Remind that the ``time''
variable in the problem is in fact the radial Liouville coordinate in a
AdS-type geometry. Hence the complex ``time'' corresponds to the
complexified radial coordinate. We shall look for a complex-time
solution which involves two Euclidean regions and one Minkowski
region between them. From quantum mechanics we know that such a
solution corresponds to tunneling through two barriers with the
classically allowed region between them. Indeed, the under-barrier
region naturally corresponds to the Euclidean region while the
interval between the barriers corresponds to the Minkowski space. We shall use
this viewpoint below. Note that in the stringy Schwinger 
production the breakdown of the minimal surface 
precisely corresponds to
emergence of a classically allowed production in the
Minkowski space \cite{gss}, supporting this general picture.

Let us start with a simplified case and tend, for a moment,
$R_{1}\to 0$ and $R_{2}\to \infty$. Then there is no scale left in
the problem and one is invited to look for a solution
\beq
\label{ansatz} r=~C\cdot z~~, 
\eeq 
where $C$ is a constant.
Substituting ansatz (\ref{ansatz}) into the equation of motion
(\ref{motion}) one finds: \beq C^{2}~=~-1/2~~, \eeq and the
respective action (\ref{action}) is pure imaginary:
\beq\label{barrier} S~=~\pm \sqrt{-1}\pi
R^{2}_{AdS}(\ln{R_{max}/R_{min}})~~, \eeq where, to regularize the
logarithmic divergence of the action, we introduced infrared and
ultraviolet cuts off. Obviously, $R_{max}\approx R_{2},
R_{min}\approx R_{1}$.

It is worth emphasizing  that the solution (\ref{ansatz}) was
recently found and discussed in Ref. \cite{alday}, in another
context. In that case  one starts with a problem  in Minkowski space
and finds that the extremal trajectory corresponds to an imaginary
action, or Euclidean signature. We start with a problem in Euclidean
space and find that the extremal trajectory corresponds to an
imaginary action, or Minkowski signature.

The actual extremal trajectory is to be more complicated than
(\ref{ansatz}). Indeed, at very small $z$ there is an ultraviolet
divergence in the action, due to the self-energy, see
(\ref{answer}). It goes without saying that we should consider
$\epsilon \ll R_{1}$. Thus, the extremal trajectory consists of
three pieces, all of which are described in terms of real  radial
coordinate $r$  and either real  or imaginary   coordinate $z$:
 \begin{eqnarray}\label{type}
 \epsilon~<~r<~R_{min}~\sim ~R_{1}:~ ~z~\approx~\sqrt{R_{1}^{2}-r^{2}}\\
 \nonumber
 R_{min}~<r~<~R_{max}~\sim ~R_{2}:~~z~\approx~ \sqrt{-2}r\\
 R_{max}<~r~<R_{2}:~~z~\approx~\sqrt{R_{2}^{2}-r^{2}}\nonumber
 \end{eqnarray}
Moreover generically one could discuss the dependence
of the radius on two coordinates similar to \cite{zarembo}.

That is the total action on the trajectory has the following
structure 
\beq S_{tot}= S_{evkl,1}(R_2)+S_{evkl,2}(R_1)
+iS_{mink}(R_1,R_2) 
\eeq 
where $S_{evkl,R_1},S_{evkl,2}$ correspond
to the actions on the cups and the imaginary contribution corresponds to
the Minkowski region. As we have seen the imaginary contribution
involves the logarithmic dependencies on the radii. Naively, the
imaginary contribution does not influence the exponential
factor which depends on $R_2$ only. Indeed in the non-conformal
case we have area law hence the area of the large loop dominates in
the action and there is no $R_1$ dependence at all. However from the
textbooks on quantum mechanics we know about the phenomena of the
resonant tunneling through  two barriers. It happens when the
energy of the incoming particle coincides with some level from the WKB 
spectrum of
the metastable states between the barriers.

Let us remind standard facts about the resonant tunneling. 
Denote the tunneling factors associated with the two barriers  as
$Z_1(E)=exp(-S_1)$ and $Z_2(E)=exp(-S_2)$. In the generic situation
the full tunneling probability is determined by the product
$Z_1\cdot Z_2$. However under the resonance condition the situation
changes, the tunneling probability reads as \beq\label{T}
T=\frac{4}{(Z_1/Z_2 +Z_2/Z_1)^2} \eeq and in case of 
symmetric barriers tends to unit. In case of the 
asymmetric barriers  the total probability depends on the ratio of two Euclidean
actions. Let us emphasize that the resonant tunneling emerges upon
taking into account an infinite number of the oscillations in the
Minkowski region.

Actually the problem we consider has a lot in common with the one
discussed in the context of multipartucle production
\cite{gv93}. Let us  remind that problem and map it to our
complex time solution. In that case one considers  decay of a
highly virtual particle into the many-body state of particles almost
at the threshold. It was argued that the problem can be reduced to
the resonant tunneling. Namely one has to look for the
solution to the classical equations of motion with the Nambu-Goto
type  action with peculiar boundary conditions. Such a solution
has been found in \cite{gv93} and can be described in 3D case as
follows. The initial  particle with energy $E$ gets transformed  into an
 expanding closed string  in the
Minkowski space. Then, at radius $r_0$, whose value  depends on the energy of the
incoming particle this string enters the Euclidean space- time region as the
solution with the same energy. The effective potential for
this problem behaves as
$$
V(R)=\sigma R
$$
where $\sigma$ is the tension of the string and R is the radius
of the circle. 

\begin{figure}
\epsfxsize=9cm
\centerline{\epsfbox{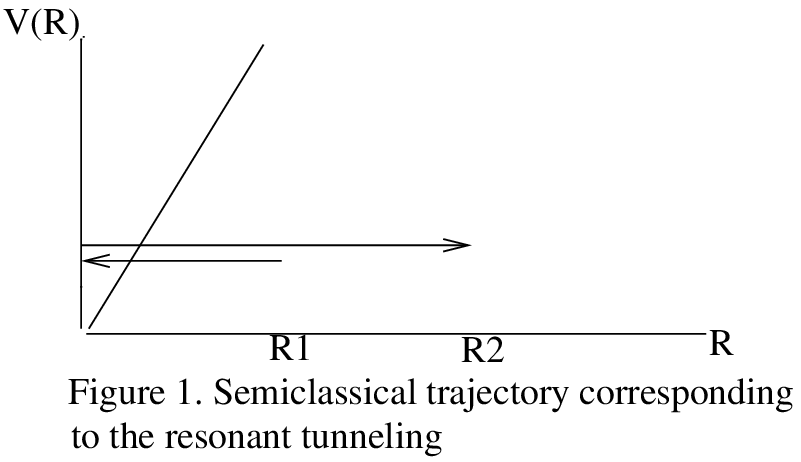}}
\end{figure}

The key point is that
this potential enjoys semiclassical levels which can be easily
calculated in the WKB approximation \beq E_N\propto
\sqrt{\sigma}N^{2/3} \eeq
and resonant tunneling happens if $E=E_N$, see Figure 1.

Let us argue that we have quite a similar situation in our problem.
Once again in the simplest case the potential for the only dynamical variable $r(z)$ is
linear and we may use it to discuss the problem qualitatively. With our
boundary conditions we look for the trajectory which starts at the small-loop
radius and upon a fixed ``time'' gets onto the  large-loop radius. The
breakdown of the minimal surface tells us  that such boundary conditions
can not be fulfilled in the Euclidean region. However they can be satisfied if
the evolution partly happens in the  Minkowski space-time
similar to the example discussed above. Generically the Minkowskian
contribution to the action is imaginary but if the resonance
condition is imposed it influences the real part as well. In our case the
resonance condition is given by the equation: \beq
\label{resonance}
S_{mink}=2\pi N\propto \sqrt{\lambda} log \frac{R_2}{R_1} 
\eeq
where N is an arbitrary integer number and 
$\lambda$ is the t'Hooft coupling.
The action calculated on the solution is proportional to the
tension of the string which  in the conformal case is proportional to the 
Yang-Mills coupling constant $g_{YM}$. Hence we could rewrite the
resonance condition in the form 
\beq 
R_{IR}/R_{UV}= exp(c/g_{YM})
\eeq 
which resembles the dimensional transmutation
formula relating the IR and UV scales in the theory with asymptotic
freedom. The coupling dependence is however different which 
could be interpreted as an indication that a nontrivial function of the coupling
constant is involved in nonconformal case which behaves as $1/g_{YM}^2$ at weak coupling
and $1/g_{YM}$ at strong coupling. In nonconformal case we can
safely identify $R_{IR}$ with $\Lambda^{-1}$.

Provided that this condition (\ref{resonance}) is fulfilled, one can
expect that there is a metastable level and the probability of the
tunneling is not actually suppressed, see (\ref{T}). This means in
turn that there is strong dependence of the correlator (\ref{one})
on the size of the small Wilson loop, or lattice spacing $a$ in case
of lattice measurements.

There are reservations to be made concerning the complex-time
solution. In particular, it is unclear if the solution we consider
is stable or not, that is if it has a negative mode in the
spectrum of fluctuations. It seems probable that it is unstable and therefore
describes the ``time'' decay of the state. Since ``time'' is our
case is identified with the radial RG coordinate then the ``decay''
would mean the RG flow from UV to IR.

\section{Minimal surfaces with the string junctions}

Generically, one can consider t'Hooft loops along
with the Wilson loops. The t'Hooft loops are obtained via the D1 string
worldsheet in the bulk  supported by a contour on the boundary. Let 
us discuss possible role of the D1 and  (p,q) dyonic strings
in our problem. The configuration
we shall look for involves the tube of the dyonic string connected
with the F1 worldsheets via junctions, see Figure 2. That is there are two
discs of the magnetic string worldsheets on two cups. To discuss the
classical configuration we have to recognize the mechanism 
of the stabilization of the radii of the discs.

\begin{figure}
\epsfxsize=9cm
\centerline{\epsfbox{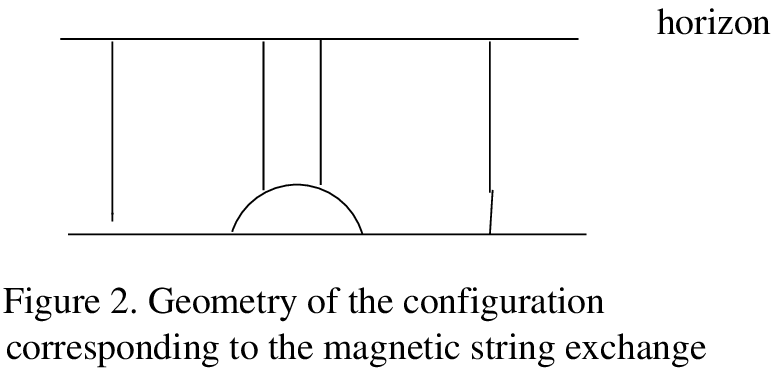}}
\end{figure}

The problem
strongly depends on the ratio of the tensions of the F1 and D1 strings.
At weak coupling F1 strings certainly are lighter than 
D1 strings. However at the strong coupling is its not necessarily true
and the opposite limit is more reliable.
Hence at weak coupling the radius stabilization  is impossible
for the dyonic tube connecting two cups, as can be seen
from the simple energy considerations.Indeed, consider the change of the total action
of the  solution due to the long tube. There are positive contributions due
to the area of the cylinder and  the 
area of the magnetic string discs since it is heavier than 
the F1 string. That is there is no reason for the stabilization
of the radius of the (1,1) cylinder and it shrinks to zero. Therefore
there is no classical solution of such type at the weak coupling.

However, a classical solution seems to exist at the strong 
coupling limit when the 
F1 string is heavier then the magnetic string. To identify the mechanism of the stabilization
let us first consider the simplest flat case when the competition of two
terms in the effective action for the cylinder radius happens: 
\beq
S_{R}= 2T_{1,1} \pi r H - (T_{(1,0)}- T_{(0,1)})\pi r^2
\eeq
where $H$ is the length of the cylinder.
That is the stabilization of the radius of the cylinder and, therefore, 
the solution to the classical equation of motion turn possible.  However,
there is a restriction on the radius of the circle since it can not
exceed the radius of the small cup $a$.  Extremizing the
effective action yields:
\beq
r_{cr}= \frac{H T_{1,1}}{\delta T}~~,
\eeq
that is we have the condition 
\beq
\frac{H T_{0,1}}{\delta T}\leq a~~.
\eeq
At the strong coupling, the length H in the nonconformal case can be approximated 
by $\Lambda_{QCD}^{-1}$
while the difference of the tensions of the strings depends on 
details of the background. Therefore we see that  the mechanism 
of the connectness of the Wilson loops correlator due to the 
exchange by the dyonic string tube can work only
in some range of fine tuned parameters.

Let us turn to a more interesting case of the AdS-type geometry. 
We are looking for a solution to the equation of motion with the real action for the
dyonic string. We expect that the dyonic string worldsheet is glued
with the cups at fixed radius and radial coordinate. That is
consider the following anzatz for the solution
\beq
(r- r_0)= C (z-z_0)
\eeq
which slightly generalizes the solution discussed above. The 
crucial point is that this modification makes the action
real, in contradistinction from   the imaginary action discussed above.
Indeed substituting this anzatz into the equation of
motion one gets
\beq
C=-\frac{z_0}{2r_0}
\eeq
and the corresponding action reads as
\beq
S_{ads}= C\sqrt{1+C^2}log \frac{R_{Min}}{R_{max}}+ ...
\eeq
Hence the solution amounts to a power-like dependence 
on the radius of the small circle. To fix the
radii one has to minimize the effective action 
with respect to $r_0, z_0$, similar to the consideration
in \cite{gss}.  The action depends on 
details of the background however the very fact of power-like dependence
of the correlator of the Wilson loops 
on the radius of the small circle is generic, 
within this mechanism of the connectness.

Let us comment on the impact of the horizon
inherent to confining backgrounds. The fundamental string
tension at the horizon is of order $\Lambda_{QCD}^2$ 
while the tension of the magnetic string effectively
vanishes. The most simple argument for this tensionless
magnetic string goes as follows \cite{gz}. Let us assume
the approach based  on the wrapped D4 branes background 
\cite{witten} and represent the magnetic string by a 
D2 brane wrapped around the angular circle of the cigar. Below the critical
temperature such wrapping is topologically unstable 
that is D2 brane tends to the tip of the cigar where
is tension clearly vanishes. This phenomena is responsible for the perimeter law of the t'Hooft loop in the confining phase. Above the critical temperature
the wrapping becomes stable and the magnetic string becomes
tensionful. That is below the critical temperature 
the magnetic string involved into the dyonic mechanism
tends to expand at the horizon region supporting the
relevance of the solution discussed above.
 
Another point to be mentioned is that the logarithmic dependence
on the small radius follows also from the deformation of the small
cup by the dyonic string tension, similar to the case discussed
in \cite{gss}. The reason for this additional logarithmic terms
is simple: they just reflect the behavior of the solution to the
Laplace equations with a source in two dimensions.

\section{Correlators involving t'Hooft loops}

Let us consider now the probe of the interior 
of the large Wilson loop by a t'Hooft or dyonic loop. That is we consider
the connected correlator
\beq
\frac{<W(R_2)W_{H}(R_1)>}{<W(R_2)><W_{H}(R_1)>}
\eeq
and discuss the mechanism of connectness via the dyonic 
string tube. Once again to discuss the classical
configuration we have to get the
stabilization of the radii of the tube.
That is the tensions of the strings with the disc topology
has to be smaller than tensions outside the tube.

Consider  the weak coupling regime when the
F1 string is light and exchange is by the dyonic string.
That is the disc at the large cup corresponds to the 
magnetic string  worldsheet  and therefore
tends to expand and its stabilization by the 
tension is possible. The disc  on the small cup can
be stabilized as well since it is made from the 
F1 string which is lighter than the magnetic string.
Hence at least at weak coupling the "`dyonic mechanism"'
is possible and it deserves detailed investigation.

The next example concerns the correlator of the large Wilson
loop and a small dyonic Wilson-t'Hooft loop
\beq
\frac{<W(R_2)W_{D}(R_1)>}{<W(R_2)><W_{D}(R_1)>}
\eeq
It is easy to see that for the (1,1) loop the mechanism
does not work.
To get the stabilization of the cylinder we have
to find the configuration when two discs at small and large cups
are the worldsheets of magnetic strings.  Taking into account the charge conservation
at the string junctions  we find, somewhat surprisingly,
that the stabilization can happen only if we consider the correlator 
of the small Wilson loop  of the (1,2) type with at least
magnetic charge two. Then the tube is the worldsheet of dyonic (1,1) 
string and the two circles are worldsheets of D1 strings. 

One has the restriction that the small radius  at equilibrium 
is to be smaller than the radius of the small loop $a$ which yields 
the restriction on the
possible set of parameters. Of course the consideration
of this section is at a qualitative level and a more
detailed analysis is required to confirm the possibility
of this mechanism. 


Finally let us consider  more exotic composite Wilson-t'Hooft loops
made from two segments, one ``Wilson segment'' and one ``t'Hooft segment''.
Since the tensions of F1 and D1 strings and their charges are different the junction
of two segments is nontrivial and has to be accomplished by an
additional (1,1) dyonic string. 

\begin{figure}
\epsfxsize=9cm
\centerline{\epsfbox{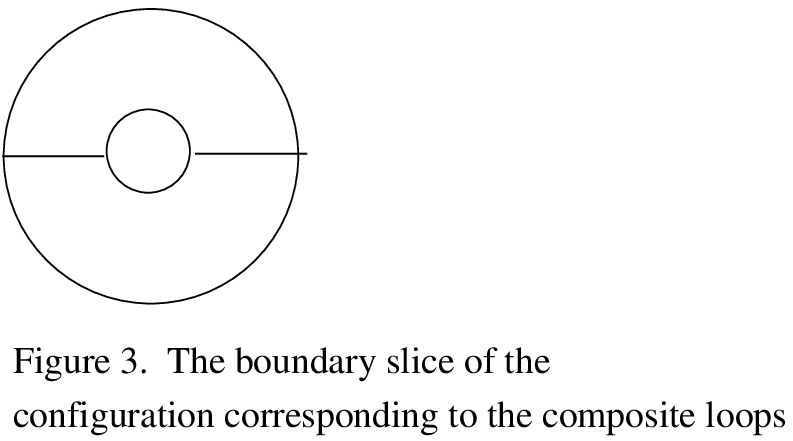}}
\end{figure}

For two parallel Wilson
and t'Hooft lines which measure the quark-monopole interaction
such worldsheet involving string junction has been considered
in \cite{minahan} where the corresponding minimal surface 
consideration has been found. Here we consider the generalization
of the similar picture for the case of two concentric composite
loops. 

 Due to the local equilibrium condition at the junction line 
and the charge conservation one has to consider the geometry
involving two ``composite cups'' and the worldsheet of the 
(1,1) dyonic string between them. The qualitative picture
of the string geometry at the boundary is presented at Fig.3. 
The action  for the configuration reads as follows
\beq
S= A_{big}+ A_{dyonic}+ A_{small}
\eeq
where three terms correspond to the area of the small,
big cups and the  area of the dyonic worldsheet between
two cups.  In the nonconformal case 
one could expect that the first term yields the ``electric'' area term while
the second yields ``magnetic'' perimeter term. Let us note that
dyonic worldsheet can be oriented differently in the
Euclidean space-time. If there is negative mode on
this composite minimal surface then these orientations yield
the different final states upon the decay process.
If there are no negative mode it is natural 
to integrate over the orientation of the dyonic string worldsheet
which  yields only inessential numerical factor.

\section{Conclusions}

In the paper we have discussed the qualitative picture behind the connected
correlator of two concentric Wilson loops in the adjoint representation with the very different radii. 
We have mainly considered three mechanisms possibly  determining intensity of
non-Abelian fields inside the flux tube. The first one is exchange
of the dilaton in bulk. It results in the prediction
\beq\label{commonsense} \langle s\rangle_0~-~\langle
s\rangle_W~\sim~(\Lambda_{QCD}\cdot a)^4~, \eeq where $a$ is the
size of the probe, or lattice spacing and the power of the factor
$(\Lambda_{QCD}\cdot a)$ is fixed by the properties of the dilaton
mode in the conformal limit.
Thus, the dilaton exchange reproduces, in the dual language, the
``common-sense'' expectation (\ref{difference}) that the fields
inside the flux tubes are soft.
 This prediction (\ref{commonsense})
does not appear self-contradictory by itself. However, it cannot
accommodate recent lattice observation \cite{boyko} that the
distribution of the fields inside the tube depends crucially on the
size of the probe.

There is another possible mechanism determining the correlator of
the two Wilson loops which is tunneling in terms of the dual
description. The extremal trajectory involves evolution both in
Euclidean and Minkowski spaces and corresponds to complex-time
solutions in quantum mechanics.
From quantum-mechanical examples it is known, furthermore, that such
solutions can produce resonant-type dependence for the correlator
on the radii of the circles.
We have a kind of parametric resonance, typical for nonlinear
systems.

Invoking this mechanism would allow to accommodate strong dependence
on the size of the small Wilson loop discovered in the lattice
measurements \cite{boyko}. However, at this moment it would be
premature to make a strong claim of 'explaining' the lattice data or
performing detailed fits of the data. Instead, we confine ourselves
to a remark that further checks are possible on the lattice. In
particular, a prediction is that dependence on the lattice spacing
is of resonance-type.

We have also qualitatively discussed the possible role of the 
configurations involving the string junctions. It was argued
that such classical configuration yields the power dependence
on the radius of the small circle. The corresponding 
power depends on the details of the nonconformal background.
We have also introduced the interesting 
objects - composite Wilson-t'Hooft loops which can not exist 
separately but there is natural ``bound state'' of two such
composite loops which certainly deserves for the further investigation.

Let us mention a few questions which could be also analyzed
within our approach. The first point concerns the temperature dependence
which strongly influences the dyonic mechanism since the magnetic
string becomes  tensionful above the phase transition point.
Another generalization concerns the Wilson and t'Hooft loops in the
different representations. The case of fundamental representation could be
of the special interest since it could be responsible for the
bound states of the different mesons. The example of such bound
state via the bulk dilaton exchange has been found recently
in \cite{dgv}. The Wilson loops in the symmetric and antisymmetric
representations are treated in terms of dielectric five-branes
and deserves for the separate analysis.

A.G. thanks INT at University of Washington 
where the part
of the work has been done during the Workshop ``From Strings to Things''
for the hospitality and support.
The work of A.G. was supported in part by grants RFBR-06-02-17382,
INTAS-1000008-7865, PICS- 07-0292165.

\end{document}